\documentclass[%
 reprint,
 amsmath,amssymb,
 aps,nofootinbib,
 superscriptaddress,
]{revtex4-2}

\usepackage{graphicx}
\usepackage{amsmath}
\usepackage{subcaption}
\usepackage{cleveref}
\usepackage{xcolor}
\usepackage{upgreek}

\newcommand{\fref}[1]{Figure~\ref{#1}}
\newcommand{\tref}[1]{Table~\ref{#1}}

\newcommand{\sref}[1]{Section~\ref{#1}}

\newcommand\blfootnote[1]{%
  \begingroup
  \renewcommand\thefootnote{}\footnote{#1}%
  \addtocounter{footnote}{-1}%
  \endgroup
}

\begin{document}

\preprint{} 

\title{Optimal Configuration of Proton Therapy Accelerators for Proton Computed Tomography RSP Resolution}

\author{Alexander T. Herrod}
\email{alexander.herrod@cockcroft.ac.uk}
\affiliation{University of Manchester, Manchester, United Kingdom}
\affiliation{The Cockcroft Institute, Daresbury, United Kingdom}

\author{Alasdair Winter}
\affiliation{University of Birmingham, Birmingham, United Kingdom}

\author{Serena Psoroulas}
\affiliation{Paul Scherrer Institut, Villigen, Switzerland}

\author{Tony Price}
\affiliation{University of Birmingham, Birmingham, United Kingdom}

\author{Hywel L. Owen\email{hywel.owen@stfc.ac.uk}}
\affiliation{STFC Daresbury Laboratory, Daresbury, United Kingdom}
\affiliation{The Cockcroft Institute, Daresbury, United Kingdom}

\author{Robert B. Appleby\email{robert.appleby@manchester.ac.uk}}
\affiliation{University of Manchester, Manchester, United Kingdom}
\affiliation{The Cockcroft Institute, Daresbury, United Kingdom}

\author{Nigel Allinson}
\affiliation{University of Lincoln, Lincoln, United Kingdom}
\author{Michela Esposito}
\affiliation{University of Lincoln, Lincoln, United Kingdom}

\date{\today}

\pacs{,,}

\begin{abstract}

	The determination of relative stopping power (RSP) via proton computed tomography (pCT) of a patient is dependent in part on the knowledge of the incoming proton kinetic energies; the uncertainty in these energies is in turn determined by the proton source -- typically a cyclotron. Here we show that reducing the incident proton beam energy spread may significantly improve RSP determination in pCT. We demonstrate that the reduction of beam energy spread from the typical 1.0\% (at 70\,MeV) down to 0.2\%, can be achieved at the proton currents needed for imaging at the Paul Scherrer Institut 230~MeV cyclotron. Through a simulated pCT imaging system, we find that this effect results in RSP resolutions as low as 0.2\% for materials such as cortical bone, up to 1\% for lung tissue. Several materials offer further improvement when the beam (residual) energy is also chosen such that the detection mechanisms used provide the optimal RSP resolution. 

\end{abstract}

\maketitle

\section{Introduction}

\blfootnote{\\\em This work is sponsored by the STFC Cockcroft Insitute Core Grant R120969/D0101}
\blfootnote{\em This work is supported partly by the EPSRC grant number EP/R023220/1}

	Proton Computed Tomography (pCT) is a technique which promises to help realise the full benefits of proton therapy \cite{bibpCTReview2015, bibpCTReview2017}, by imaging patients prior to treatment with a sparse proton beam from the same particle source used for treatment. The use of pCT improves knowledge of proton dose deposition beyond that offered by conventional X-Ray CT scans (largely due to the latter requiring the conversion from Hounsfield Units to Relative Stopping Power \cite{bibHUtoRSP}), as well as to provide imaging on the same machine as treatment is performed (potentially not having to move a patient between scanning and treatment).

	A pCT measurement typically follows the passage of a large number ($>$100 million) of individual protons through a heterogeneous region, building an image from the difference between the incident and final measured energy of each proton. Contributions to uncertainties in the measured depths and densities protons have traversed in the patient, and properties of the incident protons, increase the overall uncertainty of the imaging. Hence, these contributions should be individually considered and reduced.

	Previous studies have utilised the same energy spread as used in treatment to maximise dose rate whilst maintaining an acceptable spread of the Bragg peak.

	Here we consider for the first time how a smaller incident energy spread may benefit proton energy determination during pCT. We investigate contributions to the Relative Stopping Power (RSP) uncertainty in cortical bone, rib bone, adipose, lung tissue, water (materials relevant for head and neck imaging) and PMMA, both due to the energy spread of the incident proton beam and from the choice of incident energy.

	A separate recent study~\cite{bibDEpCT} has examined how varying the incident energy can remove imaging artefacts. Here, we propose that the RSP resolution can be optimised by reducing the incident proton beam energy spread, in combination with careful choice of incident proton energy. We start by showing -- using measurements performed at the Paul Scherrer Institut PROSCAN facility -- that a 0.2\% incident beam energy spread is reliably achievable.

	We then describe the simulations used to perform the investigations in \sref{secpCTSim}, with which both beam energy spread (discussed in \sref{secESpread}) and residual energy (in \sref{secResidE}) were optimised. The results of both optimisations are summarised in \sref{secDiscussion}, with clear improvements in RSP resolution found.

\section{Practical Demonstration of Reduced Beam Energy Spread at PSI}\label{secESpreadDemo}

	An investigation was performed\footnote{Performed as part of the INSPIRE Transnational Access (TNA) programme.} at the COMET cyclotron at PSI \cite{bibPSICyclotron}, where the energy-selection collimator aperture was reduced to provide an energy spread of 0.1\%. The energy spread of the beam was then measured at the isocentre.
	
\begin{figure}[ht!]
	    \includegraphics[scale=1.05]{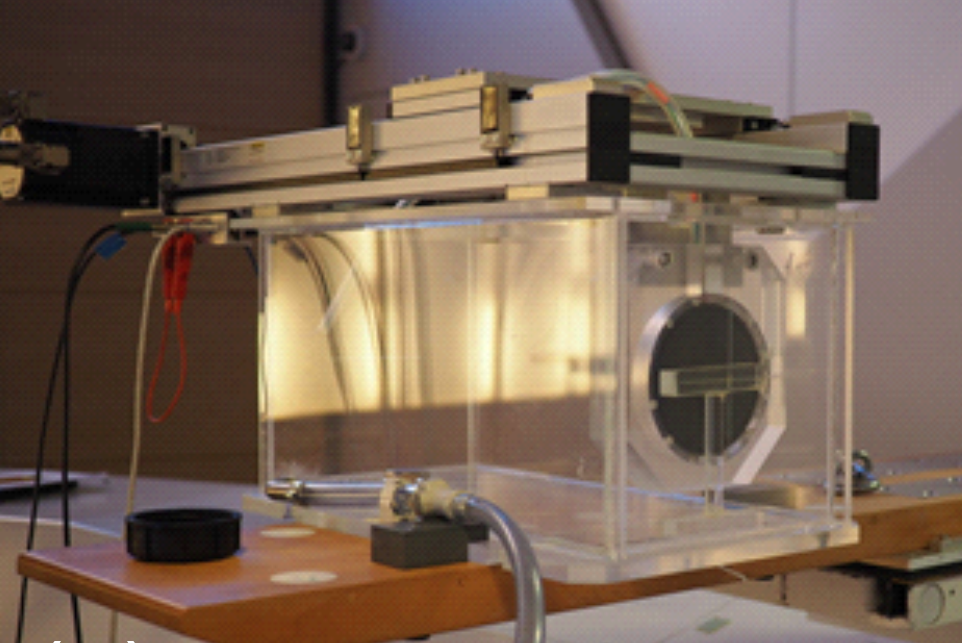}
	    \caption{\em The water range telescope used for the estimation of the minimum energy spread achievable in the clinical beam line at PSI Gantry 2. \label{figPSITelescope}}
	\end{figure}

\begin{figure}
	    \centering
	    \includegraphics[scale=0.4]{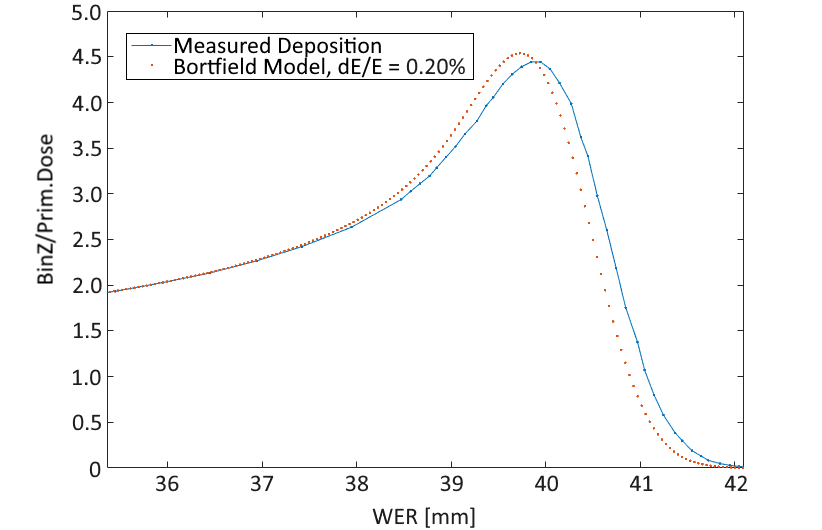}
	    \includegraphics[scale=0.4]{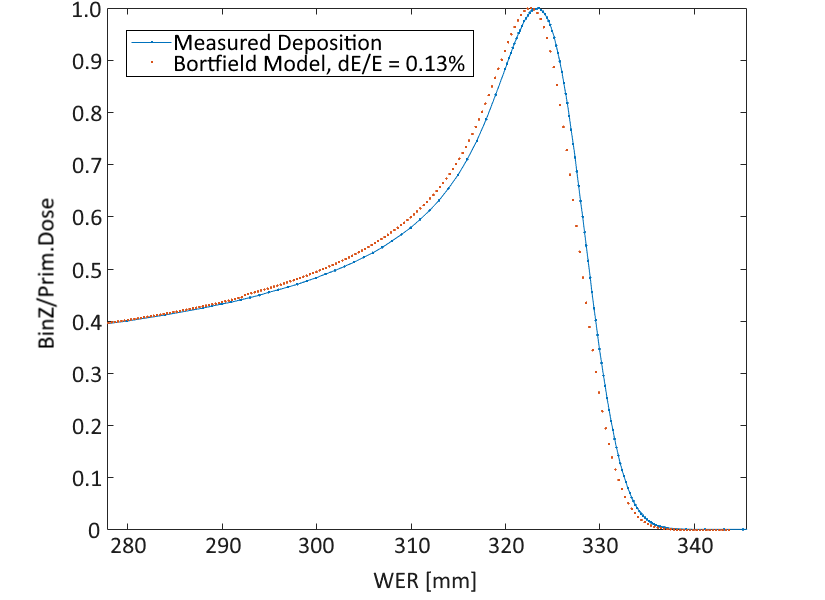}
	    \caption{\em Measured proton dose deposition in the water range telescope at PSI for minimum beam energy spread, compared with the fitted Bortfield model for 70\,MeV (top) and 230\,MeV (bottom). The widths of the Bragg peaks are fit to determine energy spread.\label{figBortfieldFit}}
	\end{figure}

	A water range telescope, consisting of a movable tracker within a water tank shown in \fref{figPSITelescope}, was used to find the position and width of the Bragg peak at beam energies of 70\,MeV and 230\,MeV, as described in \cite{Psoroulas2020}. The Bortfield model \cite{bibBortfield} was used to fit the dose deposition data from the range telescope as in \fref{figBortfieldFit}, and so find the energy spreads of the beam.
	
	It was found that beam currents between 20 and 100\,pA, sufficient for pCT scans, could be maintained with the reduction of beam energy spread to 0.2\%, for the full energy range (70-250\,MeV). Smaller energy spreads could not be achieved reliably, due to mechanical limitations in the beamline momentum spread selection collimator.
	
	Standard beam energy spreads for treatment with this machine are shown below in \tref{tabVARIANESpreads}, along with the achievable 0.2\% spreads.
	
	\begin{table}[ht!]
    	\begin{tabular}{l|c|c|c|c|c}
    		Beam Energy [MeV] & 70 & 100 & 150 & 200 & 245\\\hline
    		Beam Energy Spread & 1.1\% & 1.1\% & 0.9\% & 0.6\% & 0.3\%\\
    		Beam Energy Spread [MeV] & 0.8 & 1.1 & 1.3 & 1.2 & 0.8\\
    		$0.2\%$ of Beam Energy [MeV] & 0.1 & 0.2 & 0.3 & 0.4 & 0.5
    	\end{tabular}
    	\caption{\em Beam energy spreads of the PSI treatment beam, and the 0.2\% available on the same machine at lower currents, for e.g.~pCT scanning.\label{tabVARIANESpreads}}
	\end{table}
	
	We suggest that the reduced fractional beam energy spread at higher energy (top-right of \tref{tabVARIANESpreads}) is due to reduced use of the beam energy absorber, which spreads, as well as reduces, the proton energies.
	
	Hence, in the case where the beam is used for pCT scans, this energy spread can be reduced with a (desired) reduction in beam current, allowing for simpler proton tracking hardware and processing. We expect the most substantial improvement in resolution at lower beam energies, where the relative improvement in beam energy spread is greater. To quantify these improvements, we investigated reductions in energy spread with our simulations.

\section{Simulation of Phantom pCT Image Reconstruction}\label{secpCTSim}
	
	Proton CT scans require knowledge of the direction of and energy of protons entering and exiting an imaged object, represented here by a heterogeneous phantom. A typical pCT system will consist of a set of proton trackers both before and after the phantom to measure the incoming and outgoing proton trajectories, and an energy measurement device to measure the residual proton energy; the latter is often a range telescope based on scintillators.
	
	In this paper, we have developed a detailed GEANT4-based model (v10.0p4, physics list QGSP\_BIC\_EMY) of one potential tracking solution. We have used this to perform a virtual pCT of a phantom using a scanned pencil beam with specifications representative of those of the research beamline at the Christie Hospital proton therapy centre in Manchester (UK); based around a Varian PROBEAM cyclotron with notional fixed extraction energy of around 250~MeV, which is broadly similar to the Paul Scherrer Institut PROSCAN facility and many other high-intensity cyclotrons used for particle therapy. The GEANT4 model includes the effects of the beam, direction trackers and phantom; the resolution in the energy measurement is implemented by smearing the true residual energy of each exiting proton according to a parameterisation of the expected energy resolution as a function of proton energy.
	
	To reconstruct a pCT image of our example phantom (detailed below), we consider 360 projections in steps of 1$^{\circ}$; each projection uses 1 million protons to uniformly irradiate an 8$\times$8\,cm$^{2}$ area, giving a total of 360 million initial protons. The reconstructed trajectory and energy of the simulated protons are used to reconstruct an image using the backprojection-then-filtering method outlined in \cite{bibpCTReconstruction}.

	\vspace{-12pt}
	\subsection{Phantom}
\begin{figure}[ht!]
			\includegraphics[scale=0.15]{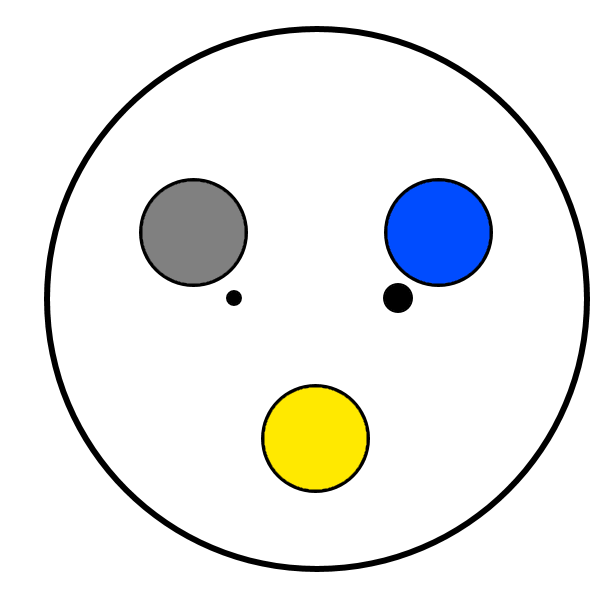}
			\includegraphics[scale=0.15]{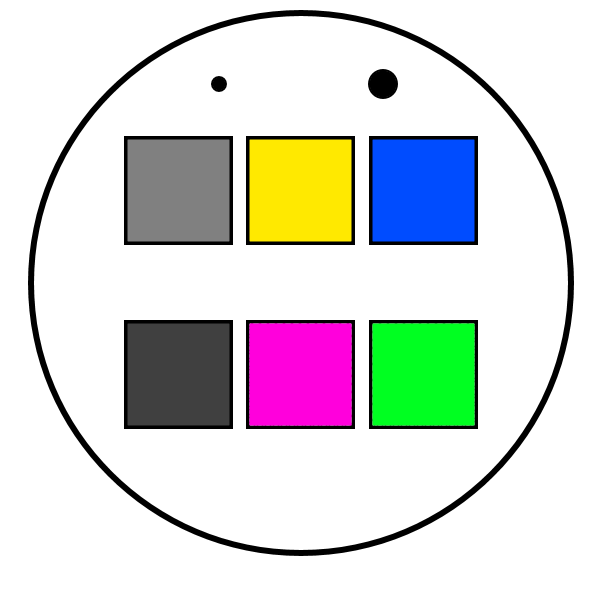}
			\caption{\em The 75\,mm-diameter phantom used in simulations, representative of the experimental PRaVDA phantom: a PMMA sphere containing six cylinders, of cortical bone (dark grey), lung (pink), air (green), rib bone (light grey), adipose (yellow), water (blue) and with two balls of tungsten carbide (black). The sphere is depicted viewed from above (left image) and from the side (right image).\label{figPhantom}}
		\end{figure}

		In simulation we replicate the experimental phantom used by the PRaVDA collaboration for earlier pCT studies \cite{bibPRaVDA}; this is a spherical PMMA phantom with radius 37.5\,mm, containing two sets of three parallel cylinders (of 7.5\,mm radius and 15\,mm length), each made of a different tissue-equivalent material: cortical bone, rib bone, adipose, water, lung and air. Two tungsten carbide balls of radius 0.5 and 1.0\,mm are also embedded at one end of the phantom. The phantom is depicted in \fref{figPhantom} and shows the materials and relative positions of the cylinders and tungsten carbide balls. 
		
		When calculating the RSP resolution for each insert, only the central region (r=2.5\,mm,Z=7.5\,mm) was considered so as to avoid spatial resolution effects that arise at material boundaries, particularly at lower energies (a method common among CT RSP studies \cite{bibpCTSimHansen,bibLMURSP}).

	\subsection{Direction Detection}
	
		The direction detection of the protons is based on using four silicon tracking modules; a pair of modules upstream of the phantom measure the incident proton trajectories, and another pair of modules downstream of the phantom measure the exist proton trajectories. Each tracking module is similar to those described in \cite{bibPRaVDATracker}, and contains 
		a 600\,$\upmu$m total thickness of silicon strips with binary readout.
		
		\begin{figure}[ht!]
			\vspace{-2px}
			\includegraphics[scale=0.55]{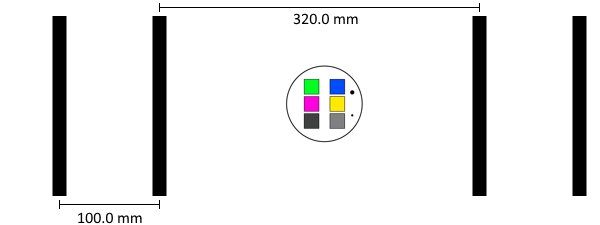}
			\caption{\em Diagram of the direction tracker configuration used. Two trackers, each consisting of 600\,$\mu$m of silicon, are situated upstream of the phantom, and two are situated downstream. The phantom sits in the centre of a 320.0\,mm separation between the centres of the inner-most trackers. \label{figTrackerDiag}}
		\end{figure}
		
		Two of these tracking modules, separated by 100\,mm, are situated upstream, and two downstream of the phantom centre, with a gap of 320\,mm between the centres of the second and third modules to accommodate phantoms of up to 220\,mm in diameter. The energy measurement device is immediately downstream of the final tracker. 
		The average position taken 
		from the planes in each tracker is used for reconstruction.
		
		The silicon layers cause some Multiple Coulomb Scattering (MCS) and energy loss within the trackers. The two most important contributions from this are: MCS from the last Si layer of the upstream module pair reduces the accuracy with which the trajectory into the phantom is known; MCS from the first Si layer of the downstream module pair reduces the accuracy with which the trajectory exiting the phantom is known. 
		
		While the tracker can be optimised to deal with multiple simultaneous protons, here we only present results based on currents where there is only ever a single proton in the system at a time, to remove any complications due to tracking ambiguities.
		
		Further effects such as strip noise, charge diffusion between strips, dead regions, detector acceptance and secondary particle production have been included in the study and are found to have minimal impact on the measured RSP resolution.

	\subsection{Energy Detection}
	
	
		Different technologies may be used for the residual energy measurement in pCT. Here, we modelled the energy measurement in two ways. In the first, we parameterise the fractional energy resolution (in percent) of the calorimeter device detailed in \cite{bibTony1} by:
		\begin{equation}\label{eqEResFit}
		    \sigma_{E\%} = \frac{146594}{E^{2.84757}} + 0.658351
		\end{equation}
		for which we present the energy resolutions in \tref{figERes}.
		
		\begin{table}[ht!]
			\begin{tabular}{l|c|c|c|c|c}
				Proton Energy [MeV] & 50 & 100 & 150 & 200 & 250 \\\hline
				Measurement Uncertainty & 2.8\% & 0.95\% & 0.75\% & 0.70\% & 0.68\% \\
				\hspace{34pt}``''\hspace{34pt} [MeV] & 1.4 & 0.95 & 1.1 & 1.4 & 1.7 \\
				WET Uncertainty [mm] & 1.1 & 1.3 & 2.1 & 3.1 & 4.4
			\end{tabular}
			\caption{\em Fractional energy measurement resolution at different energies in the range used for proton therapy, as determined from simulations. For the imaging simulations in \sref{secpCTSim}, these were approximated using the fit in \eqref{eqEResFit}. We include the corresponding Water-Equivalent Thickness (WET) uncertainty at each energy, which improves at lower energy, suggesting that the fractional energy resolution is a poor indicator of performance for pCT. \label{figERes}}
		\end{table}
		
		For the imaging simulations in \sref{secpCTSim}, the fit in \eqref{eqEResFit} was used to interpolate the energy resolution for intermediate energies.
		
		 In the other energy measurement method, we consider a hypothetical device with perfect energy measurement resolution, for which the true proton energies are used for reconstruction.


\section{Simulation Results}

	Using the energy resolutions shown in \tref{figERes} in the simulations described in \sref{secpCTSim}, we obtain the relative stopping power (RSP) values for imaging of cortical bone, adipose, rib bone, PMMA, water and lung tissue. The initial beam energies used for these simulations were 230\,MeV and 130\,MeV, with beam energy spreads of 0.2\%, and of the treatment beam, at both energies.

	Due to not all simulated protons striking the phantom, we note the mean residual energies of those which did. For the 230\,MeV beam, this mean was 200\,MeV, and was 85\,MeV for the 130\,MeV beam, suggesting a 50\% increase in average dose at the lower energy. However, we also note that the use of a spherical phantom results in a wide variation in residual energy between protons striking the phantom tangent to the surface (traversing only a few millimetres before exiting), and those striking perpendicular to the surface (traversing the full diameter).
	
	The 230\,MeV simulations produced the highest reconstruction rates at 91\% of the 360\,million simulated protons, which reduced to 81\% for 130\,MeV. This exposes our reason for not instead choosing a beam energy lower than 130\,MeV: residual protons would undergo severe scattering, which reduces the accuracy and efficiency of the tracking, and increases the path uncertainty due to multiple scattering within the phantom.

\begin{figure}[ht!]
			\includegraphics[scale=0.6]{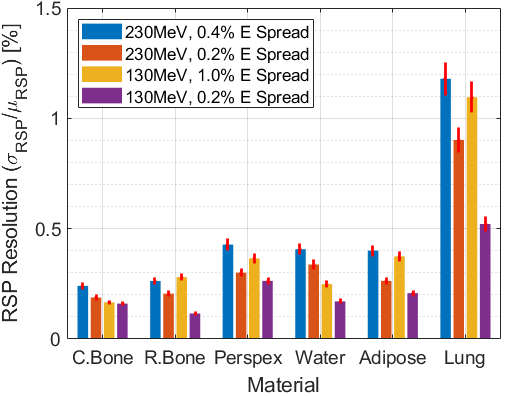}
			\caption{\em Relative Stopping Power (RSP) values determined by simulation for imaging of cortical bone, rib bone, PMMA, water, adipose and lung tissue using 130\,MeV and 230\,MeV pencil beams with energy spreads of a treatment beam (0.4\% at 230\,MeV and 1.0\% at 130\,MeV) and of 0.2\%, simulated for a perfect energy detection resolution. We see a clear improvement in RSP resolution with a reduction in beam energy spread, at both energies studied. \label{figAllRSPsUE0}}
		\end{figure}

\begin{figure}[ht!]
			\includegraphics[scale=0.6]{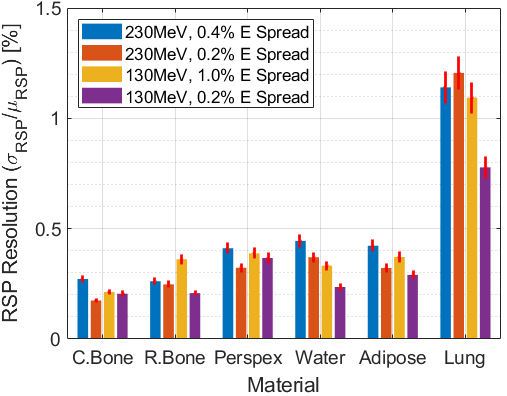}
			\caption{\em Relative Stopping Power (RSP) values determined by simulation for imaging of cortical bone, rib bone, PMMA, water, adipose and lung tissue using 130\,MeV and 230\,MeV pencil beams with energy spreads of a treatment beam (0.4\% at 230\,MeV and 1.0\% at 130\,MeV) and of 0.2\%, simulated for the residual energy detection resolutions in \tref{figERes}. We see a consistent improvement with a reduction in beam energy spread, at both beam energies studied. \label{figAllRSPs}}
		\end{figure}

	We show the resulting RSP resolutions from the simulations in \fref{figAllRSPs}, for both initial beam energies of 230 and 130\,MeV, with energy spreads of 0.2\%. We also show these results for a perfect (0\%) residual energy measurement resolution in \fref{figAllRSPsUE0}. 


\subsection{Reducing Beam Energy Spread}\label{secESpread}

		Reduction of the beam energy spread can only result in improvements for pCT, as the energies of individual protons (within the same bunch) before they enter the phantom can be known to higher accuracy. For all beam energies, the results from \sref{secESpreadDemo} indicate an available reduction in uncertainty of individual proton energy for all beam energies used, as compared to a treatment beam.
		
		By reducing the beam energy spread in the simulations to the achievable value of 0.2\%, we obtain the simulation results in \fref{figAllRSPsUE0} for a perfect energy detection resolution. Introducing the energy detection resolutions in \tref{figERes}, we obtain the results in \fref{figAllRSPs}.
		
		We see the predicted RSP resolution improvement persists for both the 130 and 230\,MeV beam energies studied. By removing the uncertainty in the residual energy measurement, we find \fref{figAllRSPsUE0} more clearly illustrates the improvement.

\subsection{Optimising Residual Energy}\label{secResidE}

	
	
	The choice of energy for a pCT scan should reflect the ideal combination of spatial resolution (not discussed or optimised here), and RSP resolution. Due to the WER uncertainty of the energy measurement improving at lower energies (\tref{figERes}), one would expect a lower energy to provide a better RSP resolution, down to energies where the worsening spatial resolution forbids study of a heterogeneous phantom.
	
	For the case of a perfect energy measurement, for beams of different energies with the same fractional energy spread (0.2\% energy spread in \fref{figAllRSPsUE0}), we find that the beam of lower energy provides an improvement in RSP resolution over the 230\,MeV beam for all materials.
	
	However, this behaviour is not as consistently observed in the results incorporating a residual energy measurement uncertainty, indicating that the energy measurement resolution (with path tracing accuracy) is critical for determining which residual energy provides the optimal RSP resolution.
	

\section{Discussion}\label{secDiscussion}
	Through assembling all results in \fref{figAllRSPsUE0} and \fref{figAllRSPs}, we see the ideal operating condition for RSP resolution in pCT is with the smallest-possible energy spread in the initial proton beam, with energy choice dependent on the tissue scanned and the method used for residual energy measurement.
	
	The reduced beam energy spread consistently produces results of between 60\% and 100\% of the RSP resolution obtained for the beam energy spread of a treatment beam, for all materials studied.
	
	While a non-zero energy measurement uncertainty serves to blur the improvement from a smaller beam energy spread, the results in \fref{figAllRSPsUE0} show this clearly.
	
	Indeed, by separating out the results for a perfect and non-perfect energy measurement (in \fref{figAllRSPsUE0} and \fref{figAllRSPs} respectively), we can decouple the energy measurement resolutions in \tref{figERes} from our results. Hence, we determine a natural improvement in RSP resolution at lower beam energies (which we associate with the increased energy absorption in the phantom, leading to a clearer indication of depth travelled), which is then counteracted by the resolution of the energy measurement method. Therefore, for other energy measurement methods, we expect the optimal beam energies to be determined by the energy resolution profile of that method.
	
	However, we note that the lower the residual energy, the higher the dose absorbed by the patient. At 130\,MeV, this dose may increase by 50\% over that at 230\,MeV. While there may be some improvement in RSP resolution for some materials at this lower energy, it is unlikely to justify the higher dose. 
	
	It is also worth noting that, although we are optimising for RSP resolution, the spatial resolution will be adversely affected by lowering the scanning energy due to increased multiple scattering within the phantom/target. The balance of RSP and spatial resolutions should be factored into treatment planning, though the reduction of beam energy spread will be of benefit at any energy.

\section{Conclusion}
	These simulations show that the RSP resolution of pCT imaging in the organic materials studied can be improved by up to 40\%, by reducing beam energy spread to an (achievable) 0.2\%. While further improvements (particularly for lung tissue) could be made by also choosing a suitable scanning/residual, this gain would be coupled with effects from spatial resolution and patient dose. Though these results were obtained for a specific scanning hardware, we expect the improvements from reduced beam energy spread to persist for all other pCT imaging technologies.
	
	Through decoupling our results from the energy measurement method (in \fref{figAllRSPsUE0}), we have found that the different materials studied exhibit between 5 and 45\% better RSP resolutions at low scanning energies (due to a higher fraction of the energy being absorbed). However, this improvement is much more consistent and pronounced (with between 45 and 65\% reduction in RSP resolution) when the energy measurement resolution is ignored.
	
	Hence, we expect a beam optimised for RSP resolution in a pCT scan to have the optimal energy for the residual direction and energy tracking to operate, and have a minimal beam energy spread, with the latter demonstrated at PSI (in \sref{secESpreadDemo}) to be readily achievable in existing machines. The latter, readily implementable change, should result (as in \fref{figAllRSPs}) in RSP resolution improvements of up to 40\% for cortical and rib bone, water, adipose and lung tissue.



\bibliographystyle{ieeetr}
\bibliography{refs}


\end{document}